\RequirePackage{fix-cm}
  \documentclass[twocolumn]{svjour3}  % twocolumn

\smartqed  % flush right qed marks, e.g. at end of proof
\usepackage{graphicx}
\usepackage[dvipsnames]{xcolor}
\usepackage{amsmath,amssymb}
\newcommand{\Ha}{{H$\alpha$}}
\newcommand{\Hb}{{H$\beta$}}

\newcommand{\Hep}{{H$\epsilon$}}

\newcommand{\Ll}{{$L_{\rm line}$}}

\newcommand{\Lacc}{{$L_{\rm acc}$}}
\newcommand{\Macc}{{$\dot{M}_{\rm acc}$}}
\newcommand{\Mout}{{$\dot{M}_{\rm out}$}}

\newcommand{\Msun}{{$M_{\odot}$}}
\newcommand{\Mjup}{{$M_{Jup}$}}
\newcommand{\Lsun}{{$L_{\odot}$}}

\newcommand{\Mstar}{{$M_{\star}$}}
\newcommand{\Lstar}{{$L_{\star}$}}
\newcommand{\Rstar}{{$R_{\star}$}}

\newcommand{\Av}{{$A_{\rm V}$}}

\newcommand{\Mdisk}{{$M_{\rm disk}$}}

\newcommand{\aap}{Astronomy \& Astrophysics}
\newcommand{\apj}{Astrophysical Journal}
\newcommand{\apjl}{Astrophysical Journal letters}
\newcommand{\aj}{Astronomical Journal}
\newcommand{\apjs}{Astrophys. Journal Suppl. Series}
\newcommand{\mnras}{M.N.R.A.S.}
\newcommand{\araa}{Annual Review of Astronomy \& Astrophysics}

\DeclareRobustCommand{\ion}[2]{%
\relax\ifmmode
\ifx\testbx\f@series
{\mathbf{#1\,\mathsc{#2}}}\else
{\mathrm{#1\,\mathsc{#2}}}\fi
\else\textup{#1\,{\mdseries\textsc{#2}}}%
\fi}

\begin{document}

\title{Accretion and Outflows in Young Stars with CUBES\thanks{Based on observations 
made with the VLT under programs 089.C-0143(A) and 097.C-0741(A)}
}
% \subtitle{CUBES observations of YSOs}


\author{J.M. Alcal\'a  \and 
    G. Cupani       \and
	C. J. Evans     \and
	M. Franchini    \and
	B. Nisini        
}


\institute{J.M. Alcal\'a \at
         INAF-Osservatorio Astronomico di Capodimonte, via Moiariello 16, 80131 Napoli, Italy \\
%        Tel.: +39-081-5575479\\
%                           Fax: +39- \\
         \email{juan.alcala@inaf.it}           %  \\
          \and
     G. Cupani \at
	 INAF-Osservatorio Astronomico di Trieste, Via G.B. Tiepolo, 11, I-34143 Trieste, Italy\\
	 IFPU--Institute for Fundamental Physics of the Universe, Via Beirut, 2, I-34151 Trieste, Italy
	   \and
	 C. J. Evans  \at
	 UK Astronomy Technology Centre, Royal Observatory, Blackford Hill, Edinburgh EH9 3HJ, UK
       \and
	 M. Franchini \at
	 INAF-Osservatorio Astronomico di Trieste, Via G.B. Tiepolo, 11, I-34143 Trieste, Italy
	    \and
	 B. Nisini    \at
     INAF-Osservatorio Astronomico di Roma, Via di Frascati 33, 00078 Monte Porzio Catone, Italy
}

\date{Received: date / Accepted: date}

\maketitle

\begin{abstract}
The science case on studies of accretion and outflows in low-mass ($<$1.5\Msun) young stellar objects (YSOs) with the
new CUBES instrument is presented. We show the need for a high-sensitivity, near-ultraviolet (NUV) spectrograph like
CUBES, with a resolving power at least four times that of X-Shooter and combined with UVES via a fibrelink for
simultaneous observations. Simulations with the CUBES exposure time calculator and the end-to-end software show that 
a significant gain in signal-to-noise can be achieved compared to current instruments, for both the spectral continuum 
and emission lines, including for relatively embedded YSOs. Our simulations also show that the low-resolution mode 
of CUBES will be able to observe much fainter YSOs (V $\sim$22\,mag) in the NUV than we can today, allowing us extend 
studies to YSOs with background-limited magnitudes. The performance of CUBES in terms  of sensitivity in 
the NUV will provide important new insights into the evolution of circumstellar disks, by studying the accretion, 
jets/winds and photo-evaporation processes, down to the low-mass brown dwarf regime.  CUBES will also open-up new 
science as it will be able to observe targets that are several magnitudes fainter  than those reachable with current 
instruments, facilitating studies of YSOs at distances  of $\sim$ kpc scale. This means a step-change in the 
field of low-mass star formation, as it will be possible to expand the science case from relatively local star-forming regions to a large swathe of distances within the Milky Way.

\end{abstract}

\section{Introduction}
\label{intro}
The study of protoplanetary disks is a rapidly growing research field. Protoplanetary disks play a pivotal role in
determining the initial conditions for planet formation, and many of their characteristics are only now being unveiled
by new observational facilities at high angular and spectral resolution, over a wide range of wavelengths, from X-ray 
to radio. 

The way in which circumstellar disks evolve and form protoplanets is strongly influenced by the processes of mass
accretion onto the star, ejection of outflows and photo-evaporation in winds of the disk material  (\cite{hartmann16},
\cite{ercolanopascucci17}). A proper understanding of the impact of these phenomena requires comprehensive  study of
different physical processes throughout the first 10$^7$ yr of the star-disk evolution. 

Observationally, the study of these processes is conducted via multi-wavelength investigations, in particular by
studying classical T Tauri stars (CTTS), which are young (a few 10$^6$ yr), very low- to solar-mass stars that are
actively accreting mass from planet-forming disks. In the current magnetospheric-accretion paradigm for CTTS, the 
strong stellar magnetic fields truncate the inner disk at a few stellar radii (\cite{DonatiLandstreet09},
\cite{johns-krull13}). Gas flows from this location onto the star along the stellar magnetic field lines, forming an
accretion shock at the stellar surface. The heated (T$\sim$10$^4$K) optically-thick, postshock gas and optically-thin,
pre-shock gas emits in the Balmer and Paschen continua and in many spectral lines (\cite{hartmann16}). At the same time,
magnetically driven winds carry away the angular momentum of the accreting gas, thereby preventing the star from
spinning up. Meanwhile, the accretion shocks produce strong UV and X-ray emission (e.g. \cite{bonito14}) that irradiates
and photo-evaporates the disk. 

Through spectroscopic surveys of young stellar objects (YSOs) in nearby (d $<$ 500pc) star-forming regions, the mutual
relationships between accretion, jets and disk structure have been addressed (e.g. \cite{alcala17}, \cite{giannini15}
\cite{frasca17} \cite{manara17} \cite{nisini18}), but further aspects remain unexplored, mainly because of the low
sensitivity and,  in some cases, the limited spectral resolution currently available in the near-ultraviolet (NUV). For instance,  
YSOs in distant  ($\sim$kpc scale) star-forming regions, where low metallicity effects  may have an important 
impact on the accretion process, have been poorly studied.

Development of the new Cassegrain U-Band Efficient Spectrograph (CUBES) for the Very Large Telescope (VLT)
\cite{zanutta22}  may enable important progress in the investigation of accretion and winds-outflows in 
solar-type young stars.

In this contribution we present the CUBES science case on accretion and wind-outflows in YSOs. Henceforth we use the
terms CTTS or YSOs to refer to solar-type young stellar objects. In Sect.~\ref{Sect2} relevant previous work and the
need for a NUV instrument with high spectral resolution and sensitivity for studies of YSOs is highlighted. In
Sect.~\ref{Sect3} the expected performances of CUBES on typical solar-type YSOs are presented, taking into account 
the possibility of a fibrelink to the Ultraviolet and Visible Echelle Spectrograph (UVES) on the VLT. For background
context, Sect.~\ref{Sect5} introduces relevant ongoing YSO projects that will influence CUBES operations, and in
Sect.~\ref{Sect4} we summarise some of the key topics in the field that CUBES will address. Finally, a summary and
conclusions are presented in Sect.~\ref{Sect6}.

\section{Previous work and the need for sensitivity and spectral resolution}
\label{Sect2}

\subsection{Measurements of accretion rates}
\label{Macc_rates}
The mass accretion rate, \Macc, can be derived from the energy released per unit time in the accretion shock 
(accretion luminosity \Lacc; see \cite{gullbring98}, \cite{hart98}) given the stellar properties, in particular 
stellar mass \Mstar\ and radius \Rstar. Observationally, 
this requires measurements of excess flux in the continuum and lines with respect to similar non-accreting template 
stars. Such measurements are best performed at ultraviolet (UV) wavelengths ($\lambda$ $<$ 4000\,\AA) 
with the Balmer continuum excess emission and the Balmer jump (see \cite{HH08}, \cite{ingleby13}, \cite{alcala14}, 
\cite{alcala17}, \cite{manara17}, and references therein). 

% In the past, \Lacc\ has been calculated using veiling measurements in high-resolution optical spectra 
% (e.g. \cite{hartigan91}, \cite{hartigan03}, \cite{white04} and references therein). 

On the other hand, accretion is a highly variable process (\cite{basribatalha90}, \cite{jayawardhana06},  
\cite{cody10}, \cite{venuti14}) which leads to a range of \Macc ~values for a given object when measured 
at different epochs  (see \cite{costigan12}, \cite{costigan14}, \cite{biazzo12}).
Variability in YSOs  induces dispersion in \Macc\ hence, in the observed \Macc--\Mstar\ and \Macc--\Mdisk\ scaling 
relationships, but cannot explain the large scatter of more than 2\,dex in $\log{}$\Macc\ at a given YSO mass. 
Such scaling relationships are predicted by the theory of viscous disk evolution (\cite{lynden-bell74}, \cite{hartmann16}, 
\cite{rosotti17} and references therein) but the \Macc--\Mdisk\ relationship has been confirmed observationally 
only recently by spectroscopic surveys in strong synergy with ALMA surveys of disks in star-forming regions 
(\cite{ansdell16}, \cite{manara16}, \cite{pascucci16}, \cite{mulders17}).

\subsubsection{The Balmer continuum} 
\begin{figure}[!ht]
\resizebox{1\hsize}{!}{ {\includegraphics[bb=-10 0 900 520]{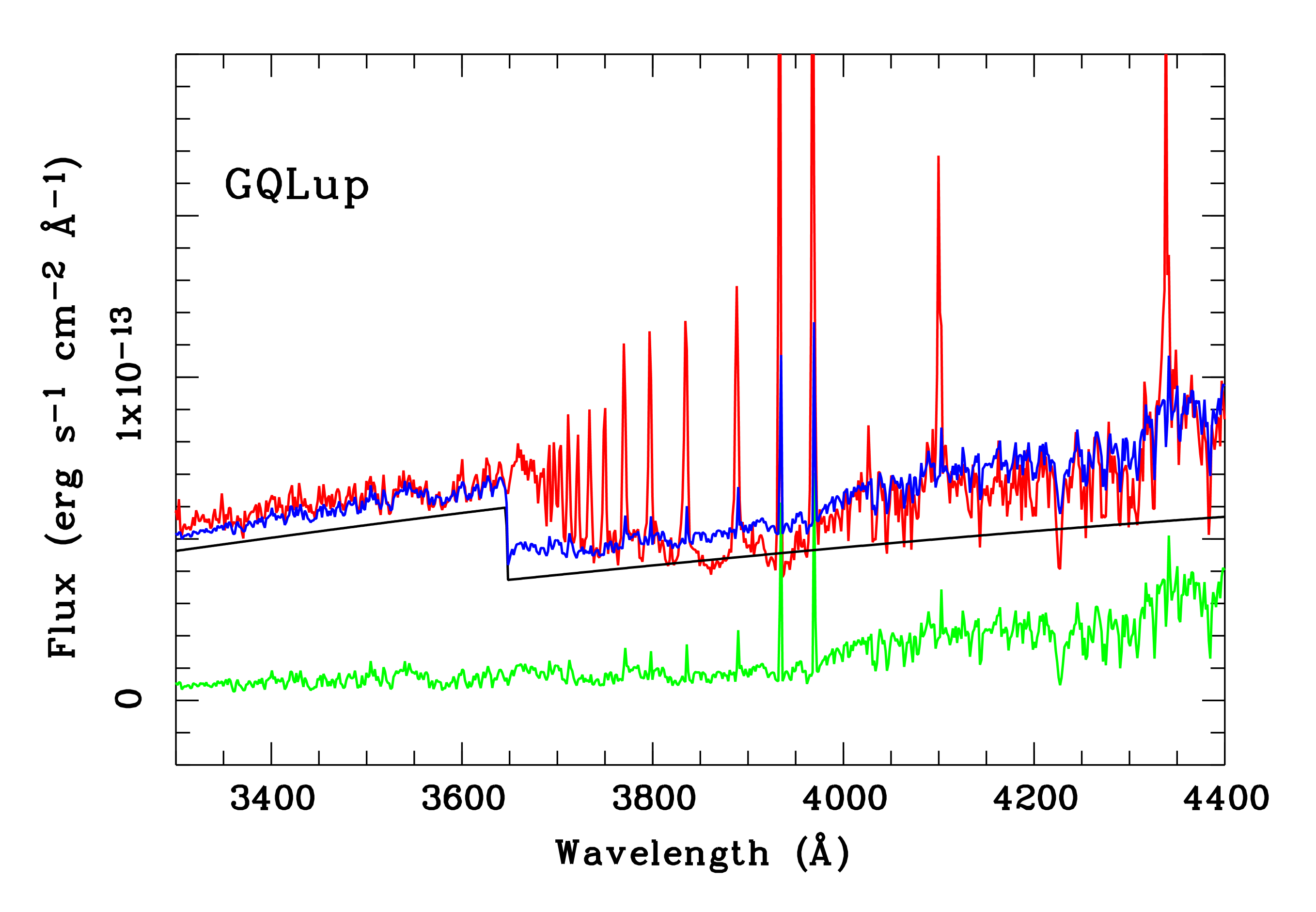}} }
\caption{Example X-Shooter spectrum of the classical T Tauri star GQ Lup in the region of the 
 Balmer jump (in red). The spectrum of a non-accreting YSO template of the same spectral type 
 as GQ\,Lup is shown in green. The continuum emission from a hydrogen slab is shown by 
 the black continuous line. The best fit to the data with the emission predicted by the 
 slab model is shown in blue. Figure adapted from \cite{alcala17}.
 \label{GQLUP_slab_fit}}
\end{figure}

The continuum excess emission in YSOs is most easily detected as Balmer continuum emission 
(see \cite{valenti93}, \cite{gullbring98} and references therein). Such continuum excess emission has 
been used in the past to derive the accretion luminosity, \Lacc, by fitting the YSO spectra with the 
sum of the photospheric spectrum of a non-accreting template and the emission of a slab of hydrogen 
(see \cite{HH08}, \cite{rigliaco12}, \cite{manara17}, \cite{alcala14}, \cite{alcala17} and references therein).
The accretion luminosity is given by the luminosity emitted by the slab. Figure~\ref{GQLUP_slab_fit} 
shows an example of the procedure in the region of the Balmer jump, applied to the X-Shooter spectrum 
of the T\,Tauri star GQ\,Lup. 

%%%%%%%%%%%%%%%%%%%%%%%%%%%%%%% Fig_ %%%%%%%%%%%%%%%%%%%%%%%%%%%%%%%%%%%%%
%%%%%%%%%%%%%%%%%%%%%%%%%%%%%%%%%%%%%%%%%%%%%%%%%%%%%%%%%%%%%%%%%%%%%%%%%%%%

In the case of X-Shooter spectra, the wide wavelength range (3000-25000\AA) allows the stellar and accretion 
parameters to be self-consistently derived by finding the best fit among a grid of slab models and 
using the continuum UV-excess emission, while constraining both the spectral type of the target and 
the interstellar extinction towards it. The best fit is found by minimizing a $\chi^2_{\rm like}$ 
distribution (see Fig.~1 in \cite{manara17a}). The procedure requires a good absolute flux calibration 
(better than $\sim$15\%) throughout the widest possible spectral range.

 High sensitivity and spectral resolution in the NUV is needed for a better definition of the Balmer jump, 
and to improve the fit to the continuum. This is particularly important in cases of low accretion rates and 
YSOs with spectral types earlier than about K3, where the contrast is low between continuum excess emission 
and the photospheric + chromospheric emission (\cite{alcala19} and see Sect.~\ref{advanced_stages}). 
A sensitivity greater  than e.g. X-Shooter is needed to detect the NUV excess emission in low-luminosity 
and slightly embedded YSOs.

\subsubsection{\Lacc--\Ll\ relationships: issues with high Balmer lines}
\begin{figure}[!ht]
\resizebox{1\hsize}{!}{ {\includegraphics[bb=0 0 350 250]{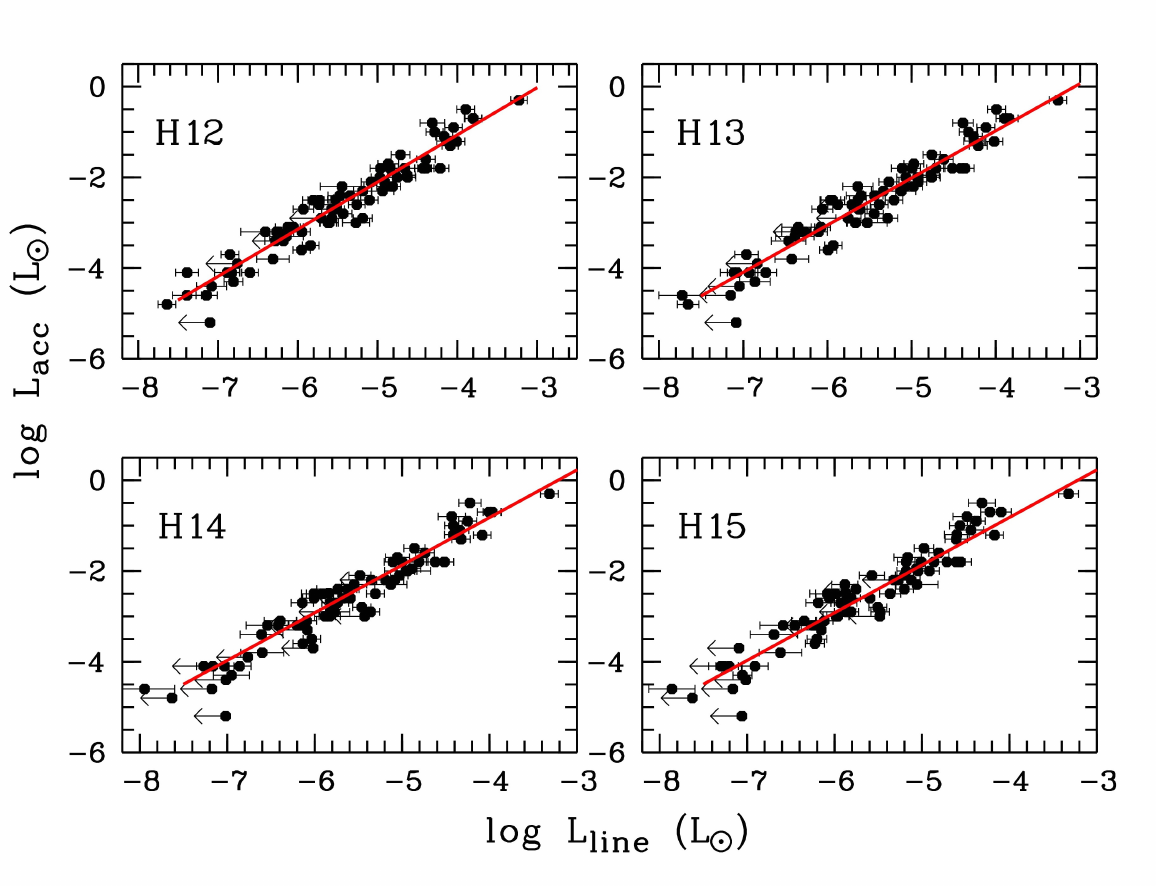}} }
 \caption{Examples of \Lacc--\Ll\ correlations for Balmer lines as labelled on each panel. 
Each point represents the \Lacc\ and \Ll\ measurement from the X-Shooter spectrum of 
an individual YSO. \Lacc\ was determined from the slab modelling as explained in the text 
(see also Fig.~\ref{GQLUP_slab_fit}), while \Ll\ is calculated from flux measurements of the 
lines and adopting a distance to the individual YSO. Leftward arrows represent upper 
limits on \Ll. The red lines represent the best linear fit to the data. Figure adapted from \cite{alcala17}.
 \label{Lacc_Ll_corr}}
\end{figure}

It is well known that \Lacc, and therefore \Macc, is correlated with the line luminosity, \Ll, of \ion{H}{i}, 
\ion{He}{i} and \ion{Ca}{ii} lines  (e.g. \cite{muzerolle98}, \cite{calvet04}, \cite{HH08}, \cite{rigliaco12}, 
\cite{alcala14}, \cite{alcala17} and references therein). These latter works provide \Lacc--\Ll\ correlations 
simultaneously and homogeneously derived from the UV to the near-infrared (NIR), underlying the importance 
of these emission features as accretion diagnostics. These accretion tracers are key diagnostics with which 
to estimate \Lacc\ via the correlations mentioned above when flux-calibrated spectra below 
$\lambda$$\sim$3700\,\AA\ are not available. Examples of these correlations, drawn from measurements 
in X-Shooter spectra, are shown in Fig.~\ref{Lacc_Ll_corr}.

%%%%%%%%%%%%%%%%%%%%%%%%%%%%%%% Fig_ %%%%%%%%%%%%%%%%%%%%%%%%%%%%%%%%%%%%%
%%%%%%%%%%%%%%%%%%%%%%%%%%%%%%%%%%%%%%%%%%%%%%%%%%%%%%%%%%%%%%%%%%%%%%%%%%%%

At low \Lacc\ values, detection of the high Balmer lines is challenging due to limited sensitivity. 
In fact, most of the lower-left points in Fig.~\ref{Lacc_Ll_corr} for the H\,15 line are represented 
by upper limits on \Ll. 
As concluded in previous works (e.g. \cite{rigliaco12}, \cite{alcala14}, \cite{alcala21}), the average \Lacc\ 
and \Macc\ derived from several diagnostics, {measured simultaneously}, has a significantly reduced error
and can also be used to simultaneously check the extinction. 
This argues for simultaneous spectroscopy from the UV to the longest practicable wavelengths, at the highest 
possible sensitivity.

Note that difficulties at the high Balmer lines may arise because of line blending, in particular 
when low low-resolution is used. Fig.~\ref{GQLUP_slab_fit} shows that the crowding of the high Balmer lines 
leads to blending that effectively shifts the Balmer limit from 3646\AA\ to the apparent jump at $\sim$3700\AA. 
Balmer lines in strongly accreting YSOs are intrinsically broad ($\Delta$V $\gtrapprox$ 250\,km/s) hence, 
the blending issues in such objects cannot be circumvented, even at the highest available resolution.
A high resolution, however, may help to both better study the line profiles and disentangle narrow 
lines superimposed on the \ion{H}{i} profile (see e.g. Fig.~\ref{Oii_Sz88A}).

\subsubsection{X-Shooter--UVES comparison}  

In this practical example we use data of the YSO Sz\,88\,A, a T Tauri star in the Lupus star-forming 
region. With an estimated \Macc$\sim$3.2$\times$10$^{-9}$\Msun/yr (\cite{alcala17}, \cite{alcala19}), this 
star is an active accretor, hence its spectrum is rich in emission lines and exhibits strong continuum 
excess emission in the UV. The interstellar extinction toward the object is low (\Av$=$0.25\,mag, \cite{alcala19}) 
and the source is sufficiently bright (V$=$13.2\,mag) for high-resolution spectroscopy with e.g. UVES. 

\begin{figure}[!ht]
\resizebox{1\hsize}{!}{ {\includegraphics[bb=5 -3 105 60]{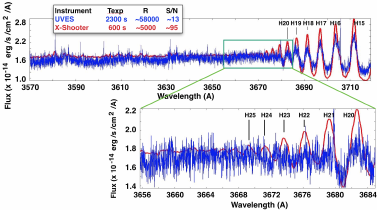}} }
\caption{Portions of the X-Shooter (red) and UVES (blue) spectra of the YSO Sz88A (large panel). 
 The upper left insert shows the corresponding exposure times, resolutions and S/N ratios. A zoom
 of the region around the Balmer jump is shown in the lower panel. Balmer
 emission lines up to \ion{H}{25} are labelled. The UVES data were normalised to the X-Shooter 
 flux at 3660\,\AA.
 \label{Sz88A} }
\end{figure}

A comparison of X-Shooter and UVES data\footnote{Unpublished data from ESO programme 089.C-0143(A), PI. B. Nisini.} 
for this object, in the spectral region around the Balmer jump, is shown in Fig.~\ref{Sz88A}. 
For comparison purposes, we have normalised the UVES data to the X-Shooter flux at 3660\,\AA.

%%%%%%%%%%%%%%%%%%%%%%%%%%%%%%% Fig_ %%%%%%%%%%%%%%%%%%%%%%%%%%%%%%%%%%%%%
%%%%%%%%%%%%%%%%%%%%%%%%%%%%%%%%%%%%%%%%%%%%%%%%%%%%%%%%%%%%%%%%%%%%%%%%%%%%

Whereas UVES provides a spectral resolution that is more than a factor of ten larger than  that from X-Shooter, the
S/N ratio  is seven times lower,  despite the much longer exposure time, even in an object as bright as Sz\,88\,A. 
Lines up to \ion{H}{25} can be detected in the X-Shooter spectrum, but they are not apparent in the UVES data
because of the low S/N. For the same reason, structure is not seen in the line profiles in the UVES data. Note,
however, that the X-Shooter spectrum was acquired more than four years before  the UVES observation. Therefore, the
non-detection of the high Balmer lines with UVES could be a consequence  of lower accretion activity in Sz\,88\,A
during the UVES observations. All this suggests the need for much  higher sensitivity in the NUV than that 
of UVES when aiming at detection of emission diagnostics during low-accretion  phases or in very weak accretors.

% On the other hand, and as diskussed above, the X-Shooter data may suffers from 
% a high level of blending at these high Balmer lines. Hence, a spectral resolution higher than that of 
% X-Shooter is necessary for the analysis of lines close to the Balmer jump.

\subsubsection{Physical conditions of the emitting gas}
\label{bal_decrs}

An important diagnostic for deriving the physical conditions (e.g. density and temperature)  of the accreting gas 
is to use the \ion{H}{i} line ratios, in particular by comparing the Balmer decrements with model predictions (e.g.
\cite{KuanFischer11}). Based on the X-Shooter data by \cite{alcala17}, \cite{antoniucci17} presented a detailed
study of the Balmer, Paschen and Brackett decrements. These authors  analysed the Balmer decrements up to the
\ion{H}{15} line, using \Hb\ ($=$\ion{H}{4}) as the reference value. 

\begin{figure}[!ht]
\resizebox{1\hsize}{!}{ {\includegraphics[bb=-10 0 900 550]{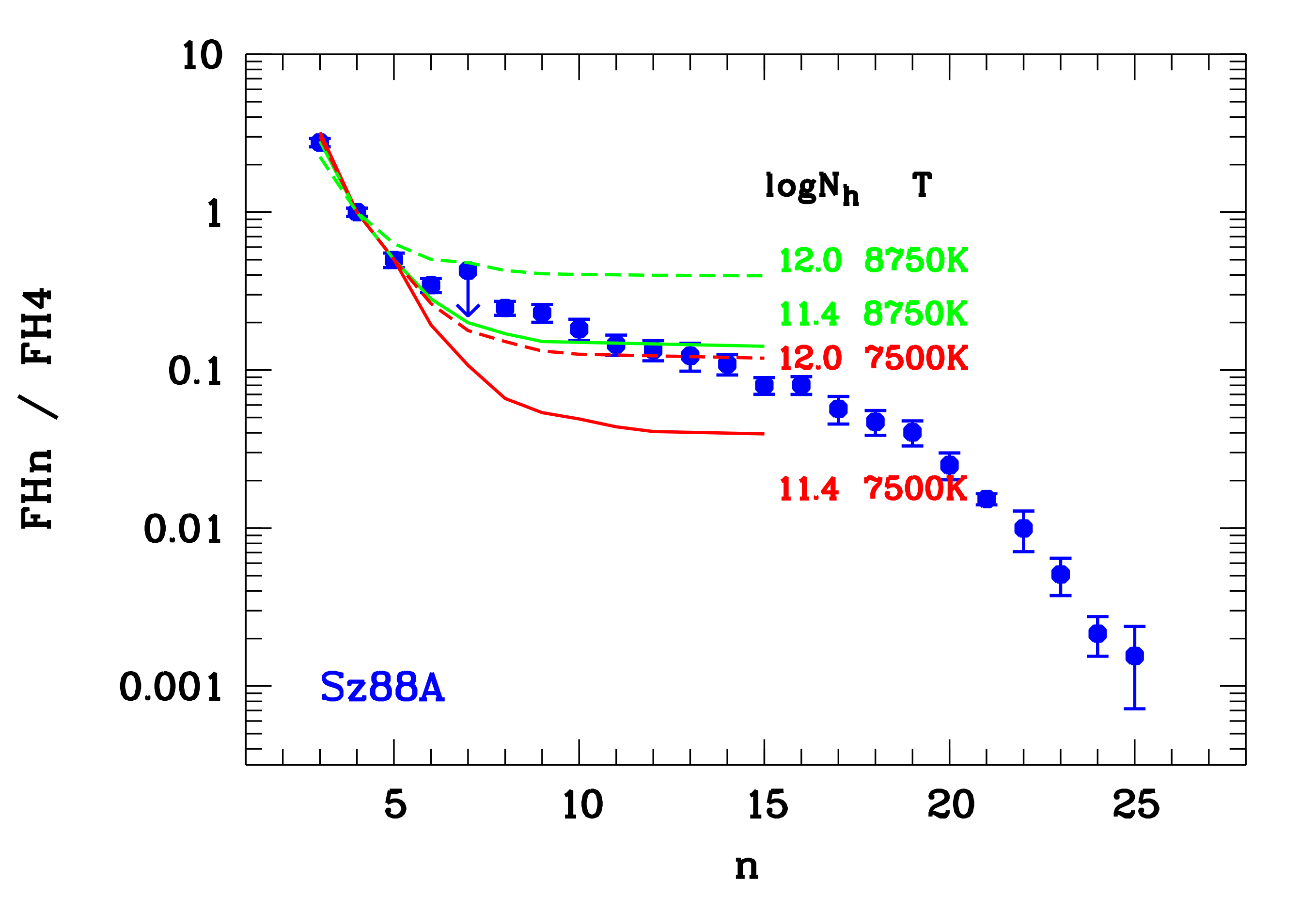}} }
\caption{Balmer decrements as a function of quantum number for the star Sz\,88\,A (blue dots), where the \Hb\ line (n=4) has been used as the reference value. The line ratio
 corresponding to \Hep\ (n$=$7) is shown as an upper limit because of problems
 of blending with the \ion{Ca}{ii}$\lambda$3969\,\AA\ line. The green and red continuous
 and dashed lines show models from \cite{KuanFischer11} with densities and
 temperatures as labelled.
 \label{bal_decr_Sz88A}}
\end{figure}

As shown in Fig.~\ref{Sz88A}, lines up to \ion{H}{25} were detected in the X-Shooter spectrum of Sz\,88\,A. 
Hence, we use these data here to exemplify the behaviour of the decrements in the high Balmer lines. 
Line fluxes were measured on the same X-Shooter spectrum, following the methodologies described 
in \cite{alcala17}. The fluxes were then corrected for interstellar extinction (\Av$=$0.25\,mag, \cite{alcala17}) 
adopting the extinction curve from \cite{fitzpatrick19}. The resulting Balmer decrements are shown in
Fig.~\ref{bal_decr_Sz88A}. For the same reasons as discussed in \cite{antoniucci17} we also used the \Hb\ 
line as the reference value. At the X-Shooter resolution, the \Hep\ ($=$\ion{H}{7}) line is not resolved
from the \ion{Ca}{ii}$\lambda$3969\,\AA\ line. 

%%%%%%%%%%%%%%%%%%%%%%%%%%%%%%% Fig_ %%%%%%%%%%%%%%%%%%%%%%%%%%%%%%%%%%%%%
%%%%%%%%%%%%%%%%%%%%%%%%%%%%%%%%%%%%%%%%%%%%%%%%%%%%%%%%%%%%%%%%%%%%%%%%%%%%

For lines up to about \ion{H}{15} the Balmer decrements in Sz\,88\,A follow the type-4, or L-shape, trend  as
defined in \cite{antoniucci17}. As demonstrated by these authors, this L-shape decrement is very well  described 
by models (\cite{KuanFischer11}) of  optically thick \ion{H}{i} emitting gas with a temperature in  the range
7500--8500\,K and densities 11.4$< \log${(N$_{\rm H}$)}$<$12  (see also the two lower-right panels in Fig.~11 of
\cite{antoniucci17}). However, the trend of the decrements for lines with quantum number higher than $\sim$15
inverts the slope, possibly indicating optically-thin  emission from gas at lower temperature. This behaviour was
also observed by \cite{rigliaco11} in the hydrogen spectrum of the brown dwarf 2MASS\,J05382543$-$0242412 (or
$\sigma$-Ori\,500) based on X-Shooter data.  The authors concluded that the  physical conditions of the emitting 
gas inferred from the hydrogen  spectrum of this brown dwarf are in contrast with the predictions of the 
magnetospheric accretion models, which predict much higher temperatures. This kind of analysis can only be 
performed for bright and highly accreting objects,  where the high-n Balmer lines can be detected.
However, the understanding of discrepancies with models requires accurate measurements of the Balmer decrement 
up to high n-number in samples of objects with different  accretion activity, which requires a greater
sensitivity than so far available and the highest possible spectral resolution.

\subsection{Wind/outflow tracers}
Mass loss through highly collimated jets and slow disk winds is a key mechanism driving accretion and contributing to
disk dissipation. Previous X-Shooter surveys of YSOs (\cite{nisini18}, \cite{nisini19}) have studied the rate of
occurrence of collimated jets, deriving a tight link with the properties of slow winds, and  establishing a
relationship between the rate of mass loss in the outflow and the mass accretion rate  (\cite{nisini19}).

While the [\ion{O}{i}] 6300\,\AA\ line has provided the historical foundation for identifying  and measuring jets and
slow disk winds, for insights into the physical conditions of the different outflow manifestations one needs to
observe several forbidden lines, probing different excitation regimes. In this respect, a key line in the NUV
spectral range is the [\ion{O}{ii}] line at 3726\,\AA, as the [\ion{O}{ii}]3726 / [\ion{O}{i}]6300 flux ratio can be
directly used to infer the degree of plasma ionisation. Knowledge of this is fundamental to infer
the total gas density (once the electron density is known from other line ratios, such as the [\ion{O}{i}]5577/6300
ratio)  and thus to have a correct measurement of the mass-loss rate, which is a critical parameter to constrain the
different models for disk dissipation  and thus understand the protoplanetary disk evolution and planet formation
(e.g. \cite{Kunitomo2020})

\begin{figure}[!ht]
\resizebox{1\hsize}{!}{ {\includegraphics[bb=3 -1 54 35]{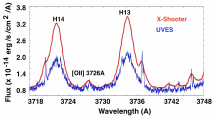}} }
\caption{Portions of the X-Shooter (red) and UVES (blue) spectra of Sz\,88\,A in a
 spectral range around the [\ion{O}{ii}] 3726\,\AA\ line. The spectra are the same as
 in Fig.~\ref{Sz88A}. The UVES data were normalised to the X-Shooter 
 flux at 3660\,\AA.
\label{Oii_Sz88A}}
\end{figure}

Gas ionization is low in T Tauri outflows (e.g. a fraction of 0.1-0.3) so the [\ion{O}{ii}]~3726\,\AA\ line is usually weak and therefore requires
sensitive NUV observations to detect it in different sources. In addition, and as illustrated in
Fig.~\ref{Oii_Sz88A}, a spectral resolution higher than that provided by X-shooter is also needed to resolve the 
line into components at different velocities and thus measure the ionization fraction in each of them. This example
effectively shows that the [\ion{O}{ii}] can be resolved into two components, possibly  of low and high velocity,
that cannot be resolved at the X-Shooter resolution. We note that  the [\ion{O}{ii}] is a doublet at
3726.0/3728.8\,\AA\ and what we are seeing in Fig.~\ref{Oii_Sz88A} are two velocity components of  the
[\ion{O}{ii}]3726.0\,\AA\ line, separated by 0.7\,\AA\ i.e. about 56 km/s which is a typical velocity  separation in
T Tauri outflows. The other component of the doublet at 3728.8\,\AA\ (at  2.8\,\AA\  from the  3726.0\,\AA\ line) is
barely visible above the noise in the UVES spectrum, as expected having a radiative  rate a factor of 10 smaller 
than the 3726.0\,\AA\ line.

Notably, the intensity of the forbidden line in Fig.~\ref{Oii_Sz88A} remains approximately the same, whereas the
intensity  of permitted lines, such as \ion{H}{13} and \ion{H}{14}, is clearly changing. As mentioned before, the
latter may be a consequence of variable accretion, as the UVES spectrum was acquired more than four years after 
the X-Shooter observation. 

%%%%%%%%%%%%%%%%%%%%%%%%%%%%%%% Fig_ %%%%%%%%%%%%%%%%%%%%%%%%%%%%%%%%%%%%%
%%%%%%%%%%%%%%%%%%%%%%%%%%%%%%%%%%%%%%%%%%%%%%%%%%%%%%%%%%%%%%%%%%%%%%%%%%%%

% This example effectively shows that the [\ion{O}{ii}] can be % resolved in two components, possibly 
% of low and high velocity, that cannot be resolved at the 
% X-Shooter resolution. 

\section{CUBES performance for YSO studies}
\label{Sect3}
The current topics in YSO studies and limits of observations outlined in Sect.~\ref{Sect2} argue for a
high-sensitivity spectrograph such as CUBES, with a resolving power at least four times that of X-Shooter, 
and possibly linked to UVES for simultaneous observations with both instruments. We now present results 
from simulations to quantify the potential performance of CUBES in the context of YSO studies.

The primary operational mode of CUBES is expected to have a resolving power R$\geq$20,000 over the spectral range
from 3000\,\AA\ to 4050\,\AA. Taking advantage of the high-sensitivity provided by CUBES in the near-UV, a second
operational mode with R$\approx$7,000 over the same spectral range is foreseen to allow the investigation of
background-limited objects for science cases not requiring the high-resolution mode. In addition, a fibrelink 
to UVES for simultaneous observations at longer wavelengths is included in the conceptual design \cite{zanutta22}.
 
To investigate the instrument performance for the YSO cases we used the exposure time calculator 
(ETC)\footnote{http://archives.ia2.inaf.it/cubes/\#/etc} and the end-to-end simulator (E2E)
software\footnote{https://cubes.inaf.it/end-to-end-simulator} developed for CUBES \cite{genoni21}. To apply these
tools to our science case, accurately flux-calibrated spectral templates of  YSOs are needed, with good S/N ($>$90)
in the NUV, and with a resolution similar or  higher than that expected for CUBES. Real spectra, simultaneously
satisfying such characteristics,  are not available, while reliable models of spectra reproducing the complete YSOs
phenomenology do not  yet exist. Real data that most closely satisfy these requirements are the X-Shooter  spectra.
Despite their lower resolution, the general spectral shape can be used to predict the S/N of  the continuum (and the
higher S/N expected for the emission lines). We therefore used these as our input templates, in particular those of
the YSOs discussed in Sect.~\ref{Sect2}, namely GQ\,Lup and Sz\,88\,A. A shortcoming of the X-Shooter data is that
the S/N drops significantly for $\lambda$ $<$ 3250\,\AA\ hence, for simplicity, we set the flux to zero below this
wavelength. 
%{\it In any case, excess emission is also expected in the short wavelength range and the S/N will not be very different as in the spectral regions around 3600\,\AA, depending on the slope of the Balmer continuum.} \textcolor{magenta}{I don't think this necessarily follows given the significant increase in atmospheric absorption at shoter wavelengths? Could omit last sentence?}

\subsection{High-resolution mode}
\label{high_res_mode}

\paragraph{\underline{Brighter objects (V$<$19\,mag):}}
We used the X-Shooter observation of GQ\,Lup (Sz\,75; SpT$=$K6,  \cite{alcala19}) as our input template, assuming
\Av$=$0\,mag. We used the CUBES ETC to estimate the performance for V$=$10\,mag and 19\,mag, in adopted conditions
of  seeing$=$1\,arcsec and airmass=1.2, with predicted S/N ratios (at $\lambda$=3600\,\AA) of $\sim$320 and  10, in
5\,min and 1\,hr, respectively.  For an input template with a weak continuum Balmer emission, and with an earlier
spectral type than GQ\,Lup (e.g. MY\,Lup, SpT$=$K0, \cite{alcala19}), slightly lower values of S/N are predicted,
with S/N = 300 in 5\,min and 9 in 1\,hr for V$=$10\,mag and 19\,mag, respectively. Adopting V$=$18.5\,mag (for the
MY\,Lup template) we find S/N$=$14 in 1\,hr;  this magnitude is typical of YSOs in the low-metallicity region Sh2-284
\cite{cusano11}.

%%%%%%%%%%%%%%%%%%%%%%%%%%%%%%% Fig_ %%%%%%%%%%%%%%%%%%%%%%%%%%%%%%%%%%%%%
\begin{figure*}[!ht]
\resizebox{1.0\hsize}{!}{ {\includegraphics[bb=5 0 110 60]{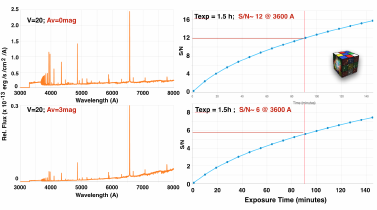}} }
\caption{Results of the CUBES ETC for the faint case discussed in the text. The template 
spectra for visual extinctions of \Av$=$0 and 3\,mag are shown in the 
upper and lower left-hand panels, respectively. Note that the unreddened flux of the template 
corresponds to the actual magnitude of Sz\,88\,A (V$=$13.2\,mag), but it is internally 
normalised to V$=$20\,mag in the ETC. The ETC results for 1.5hr exposures for the two \Av\ cases are shown in the right-hand panels. 
The blue curves with dots in the right-hand panels show the S/N ratio as a function of 
exposure time for $\lambda=$3600\,\AA. The resulting S/N ratios at 3600\,\AA\ are labelled 
in red. All the plots were extracted from the ETC.
\label{Sz88A_ETC_result1}}
\end{figure*}
%%%%%%%%%%%%%%%%%%%%%%%%%%%%%%%%%%%%%%%%%%%%%%%%%%%%%%%%%%%%%%%%%%%%%%%%%%%%

Adopting a later M6 template (2MASS\,J16095628$-$ 3859518 or Lup818s in \cite{alcala14}) with V$=$19\,mag and 
the same conditions in the ETC yielded S/N$\approx$10 in 1\,hr. This example highlights the interesting prospect 
of systematically studying accreting brown dwarfs at relatively high resolution in the UV for the first time.

\paragraph{\underline{Faint objects (V$>$19\,mag):}}
For objects as faint as V$=$ 20\,mag, we use the X-Shooter data of Sz\,88\,A as the input template,
which also has a very good S/N and was used to exemplify the behaviour of the Balmer decrement
in Sect.~\ref{bal_decrs}. 
This template is shown in the left-hand panels of Fig.~\ref{Sz88A_ETC_result1} for the cases of \Av$=$0 and 
3\,mag of visual extinction.
Using the CUBES ETC and adopting conditions of seeing$=$0.8\,arcsec and airmass=1.2, we estimate 
a S/N of 12 and 6 at $\lambda=$3600\,\AA\ for the two \Av\ values, respectively, 
in 1.5\,hrs of exposure and with binning $\times$2 in the spatial direction. 
The S/N ratio at 3600\,\AA\ as a function of exposure time is shown in the right-hand panels of  
Fig.~\ref{Sz88A_ETC_result1} for the two \Av\ values. We also used the E2E simulator 
with the same template and observing assumptions and recovered an estimated S/N that was in very good 
agreement with the ETC.

For comparison with the above results, 
we obtained 
%we used the {\em splot } task in IRAF\footnote{IRAF is distributed by the National Optical Astronomy Observatory, which is operated by the Association of the Universities for Research in Astronomy, inc. (AURA) under cooperative agreement with the National Science Foundation.} to measure a
S/N$\approx$5 at $\sim$3600\,\AA\ 
in our X-Shooter spectrum of the young brown dwarf 2MASS\,J05382543$-$ 0242412 
($\sigma$-Ori\,500 with V$=$20\,mag, \Av$=$0\,mag, \cite{rigliaco11}). 
This 1.5\,hr exposure was done under similar conditions as those assumed for the ETC predictions 
above, meaning that in the faint example (with \Av$=$0) CUBES would provide data with twice the 
S/N and about four times better resolution than X-Shooter. 
In addition, the quality of the CUBES data for moderately embedded objects will be sufficient for 
the studies discussed in Sec.~\ref{Sect4}.

It is worth mentioning that the S/N ratio per pixel provided by the ETC does not 
take into account any optimal extraction of the 1-dimensional spectra and therefore the predicted 
S/N value is expected to improve in case of optimal extraction from the 2-dimensional images. 
Also, the S/N of the emission lines will be higher than for the continuum.

%%%%%%%%%%%%%%%%%%%%%%%%%%%%%%% Fig_ %%%%%%%%%%%%%%%%%%%%%%%%%%%%%%%%%%%%%
\begin{figure*}[!ht]
\resizebox{1.0\hsize}{!}{ {\includegraphics[bb=2 0 90 62]{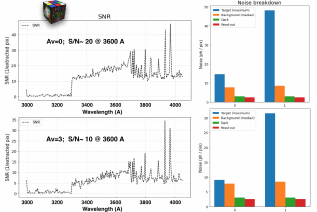}} }
\caption{Results from the CUBES E2E simulator for the low-resolution mode and the faint case (V$=$20\,mag) 
discussed in the text. The left-hand panels show the resulting S/N as a function of wavelength for visual extinctions \Av$=$0 and 3\,mag as indicated. Note the higher S/N of the emission lines 
with respect to the continuum and the zero S/N for $\lambda<$3250\,\AA\ due to the cut of flux
in the input X-Shooter data. The right-hand panels show the noise breakdown for the two corresponding \Av\ 
values. The two histograms in each panel are the photons/pixel of each component as coloured and
labelled in the legend, for the two instrument arms (0 or 1 on the x-axis, the `blue' and `red' arms, respectively). Plots extracted from the E2E simulator.
\label{low_res_e2e_1}}
\end{figure*}
%%%%%%%%%%%%%%%%%%%%%%%%%%%%%%%%%%%%%%%%%%%%%%%%%%%%%%%%%%%%%%%%%%%%%%%%%%%%
 
\subsection{Low-resolution mode}
\label{low_res_mode}

We make use of the E2E software to simulate the low-resolution mode of CUBES, adopting the first configuration 
described in Sect.~4.2 of \cite{genoni21}, which uses six slices of 1\,arcsec each.  With the same input parameters
as in the previous  subsection (Sz\,88\,A template, V$=$20\,mag, seeing $=$ 0.8 arcsec, airmass$=$1.2,  T$_{\rm
exp}=$ 1.5\,hr, $\lambda=$3600\,\AA)  we obtained the results shown in Fig.~\ref{low_res_e2e_1}, namely S/N values of
20 and 10 for the \Av$=$0 and 3\,mag  models, respectively, with a spatial binning $\times$4 and wavelength binning
$\times$3 (which still yields a sampling well above the Nyquist limit). The noise breakdown histograms are also shown
in the right-hand panels  of the figure. Note the limited contribution of the target on the spectrograph arm `0',
which is due to the zero  flux at $\lambda<$3250\,\AA\ in the input template.

\begin{figure*}[!ht]
\resizebox{1.0\hsize}{!}{ {\includegraphics[bb=-5 0 68 60]{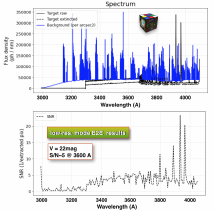}} }
\caption{Upper panel: the E2E input spectrum of Sz\,88\,A with and without taking into account airmass (black dashed and continuous lines, respectively) for the case of a V$=$22\,mag YSO and \Av$=$0\,mag.; the background contribution is shown 
with the blue line. Lower panel: the E2E output S/N as a function of wavelength (dashed line) for 
a 2 hour exposure. Note the zero S/N for $\lambda<$3250\,\AA\ due to the cut of flux of the input X-Shooter 
data. Plots extracted from the E2E simulator.
\label{low_res_e2e_2}}
\end{figure*}

This simple exercise shows that a significant gain in S/N, both in continuum and lines, can be achieved 
with the CUBES low-res. mode, and suggests that observations of even fainter YSOs can be attempted if we are 
able to trade spectral resolution for sensitivity. Hence, pushing to lower flux limits, for V$=$22\,mag and 
\Av$=$0\,mag, with T$_{\rm exp}=$2\,hr and adopting the same parameters as before, the E2E simulator yields 
the results shown in Fig.~\ref{low_res_e2e_2}. 
The predicted S/N is $\sim$5 at $\lambda=$3600\,\AA, but it is $>$10 for many emission lines, which
opens-up the possibility to extend studies done with X-Shooter to YSOs with 
background-limited magnitudes.

We note that the result from the E2E simulator is the combination of the extracted spectra from all 
the instrument slices (see \cite{genoni21}); in our simulations we only use three slices but most of the object's 
flux will go through the central slice, hence we expect an optimal extraction of this slice would further 
improve the final S/N, compared to the estimates shown here.

%%%%%%%%%%%%%%%%%%%%%%%%%%%%%%% Fig_ %%%%%%%%%%%%%%%%%%%%%%%%%%%%%%%%%%%%%
%%%%%%%%%%%%%%%%%%%%%%%%%%%%%%%%%%%%%%%%%%%%%%%%%%%%%%%%%%%%%%%%%%%%%%%%%%%%

\subsection{Link with UVES}
\label{uves_link}

The simultaneous fit of the Balmer continuum and physical parameters of YSOs is best when the widest possible
spectral window is available, which requires observations that extend  redwards far beyond the CUBES range (see
Sect.~\ref{Sect2}). A fibrelink to UVES for observations with both instruments simultaneously would make this
possible. As accretion is a highly variable process, the simultaneous link is necessary for the fitting procedure
described in Sect.~\ref{Sect2}. 

Introducing a fibrelink entails a reduction in the UVES throughput, so this is only an efficient option if the
performance of UVES with the fibrelink is sufficiently good to deliver the required observations within the time
needed for CUBES observations. The conceptual design provides a 40\% throughput for the fibre-link at $>$5000\,\AA. 
We estimated the exposure times required to reach a S/N=10, adopting the unreddened spectral template shown in 
Fig.~\ref{Sz88A_ETC_result1} (in the reddened case, the NUV performance will be the  more limiting factor); for
airmass$=$1.2 and seeing$=$0.8 arcsec for:

\begin{itemize}
\item[$\bullet$] CUBES ETC: continuum near the [\ion{O}{ii}] 3726\,\AA\ line, V$=$19.5\,mag, with the high-resolution  
     option and binning $\times$2 spatially, gives S/N$\approx$10 in 1\,hr.
\end{itemize}

\begin{itemize}
\item[$\bullet$] UVES ETC: continuum near the [\ion{O}{i}] 6300\,\AA\ line, V$=$20.5\,mag (scaled to mimic the reduced 
     throughput of the fibrelink compared to normal UVES operations), with a 1.0" slit, and the red-arm 
     (580\,nm) setting. The latter combination gives a resolution R$\approx$40,000 with sampling of $>$5 pixels. 
     With 2$\times$2 binning we recover a S/N$\approx$5.5 in 1\,hr. Binning this by a further factor of three 
     (oversampling, degrading resolution to match CUBES), gives a comparable S/N in a similar exposure time; 
     i.e. the UVES exposure time fits within the CUBES observation.
\end{itemize}

\noindent
Without the fibrelink, a red-arm UVES exposure of $\sim$1200s gives comparable S/N (ignoring the preset and 
separate acquisition time), i.e. each combined (CUBES + UVES) observation
effectively saves half an hour 
of otherwise separate UVES observations in this example.

\section{Synergies with other projects}
\label{Sect5}
For background context of future programmes in this field, we now briefly introduce some on-going large programmes on YSOs that will have strong synergies with CUBES observations.

\subsection{ALMA disk surveys}
\label{disk_surveys}
The Atacama Large Millimeter Array (ALMA) provides sufficient sensitivity and resolution at 
sub-mm wavelengths to detect and measure the mass, \Mdisk, of dusty protoplanetary disks around YSOs with 
a mass down to 0.1\,\Msun\  (\cite{ansdell16},  \cite{ansdell18}, \cite{pascucci16}, \cite{barenfeld16},
\cite{cazzoletti19}). 
Such surveys have provided the demographics of disks in nearby star-forming regions. The synergy  of these surveys
with the studies of the properties of the central objects, such as stellar mass, \Mstar,  and accretion rate, \Macc,
have provided important constraints on models of viscously evolving disks by confirming  the correlations of \Mdisk\ 
with \Mstar\  and \Macc\ (see Sect.~\ref{low_end_of_Macc}), although with significant  scatter. However, these
synergies did not include statistically significant samples of YSOs with \Mstar$\le$ 0.1\Msun,  preventing
investigation of whether the scaling relationships are different for low-mass substellar  objects to those for stars.
Future sensitive ALMA surveys of compact and small disks in very low-mass YSOs,  will allow investigation of the
correlations down to the low-mass brown dwarf regime.

\subsection{ULLYSES \& ODYSSEUS}

The public {\em Hubble Space Telescope} UV Legacy Library of Young Stars as Essential Standards (ULLYSES) survey will
enable significant advances in the study of astrophysical disk accretion processes. A total of 500 orbits of this
Directors Discretionary Time programme has been dedicated to far-UV spectroscopy with the Cosmic Origins Spectrograph
(at R$\approx$18,000) and low-resolution NUV/optical spectroscopy with the Space Telescope Imaging Spectrograph of 
tens of solar-type YSOs. The targets cover a range in stellar mass, accretion rate, and age  (e.g., \cite{alcala17},
\cite{manara17}). 

A large, international team has initiated the Outflows and disks around Young Stars: Synergies for the Exploration of
ULLYSES Spectra (ODYSSEUS) project to bring together  the broad expertise required to resolve the fundamental
problems on accretion and wind on YSOs through the investigations enabled by ULLYSES \cite{espaillat22}, and to
coordinate the ambitious datasets of simultaneous and  contemporaneous observations that will enhance the impact of
the public data (e.g. PENELLOPE \cite{manara21}).  We anticipate strong synergies of future CUBES observations of
YSOs with ODYSSEUS science products.

\subsection{PENELLOPE}
The ESO Large Programme (106.20Z8) PENELLOPE (\cite{manara21}, \cite{frasca21}) at the VLT is providing contemporaneous
high-resolution (UVES, ESPRESSO) and UV-NIR mid-- resolution (X-Shooter) spectra of the ULLYSES targets to provide
their accretion/wind and stellar physical and kinematical properties.  The project involves more than 50 researchers
and, together with ODYSSEUS, will provide well characterized targets in terms of accretion/wind and stellar
properties. These will then serve as a bright benchmark sample for CUBES programmes. 

\subsection{GHOsT}
The GIARPS High-resolution Observations of T Tauri stars (GHOsT) (\cite{giannini19}, \cite{gangi20},
\cite{alcala21})  is a GIARPS@TNG optical/IR high-resolution spectroscopic survey of a flux-limited, complete sample
of T Tauri stars in  the Taurus star-forming region. This ongoing survey will homogeneously derive (i.e. avoiding
systematics due to non-simultaneous observations) stellar and accretion/outflow parameters of sources in Taurus, and
constrain the properties of the gas in the inner disk regions. This survey is in synergy with existing ALMA disk
observations in the region (e.g. \cite{long19}), so the programme will directly link the processes occurring in the
star-disk interaction region  and the overall disk structure, providing a unique, rich, and robust observational
reference for disk evolutionary models.

\section{Studies of YSOs with CUBES}
\label{Sect4}
The high performance of CUBES in terms of spectral resolving power and sensitivity will  mostly broaden 
our knowledge on accretion and outflows in YSOs, but will also open-up new science as it will be able to observe 
targets that cannot be reached with current facilities. In this section we describe some of the YSO topics that 
can be addressed with CUBES, and with the proposed link to UVES.  Except for line profile analysis 
(line shapes and peak position, widths and kinematics), these studies require a good absolute flux calibration 
($\sim$15\% or better, see Sect.~\ref{Macc_rates}) throughout the widest spectral range. If this requirement cannot be satisfied for simultaneous CUBES + UVES observations, contemporaneous photometry will be needed, from the NUV to the reddest possible wavelength.

\subsection{Accretion}
High throughput observations at R$\ge$20,000 will enable more detailed study of the accretion process 
than currently possible, via precise modelling of the Balmer jump, and by studying the higher-order 
Balmer lines. The high Balmer and \ion{Ca}{ii} H (3968\,\AA) \& K (3933\,\AA) lines provide diagnostics of the 
accretion funnel flows and heated chromosphere in the post-accretion shock region (e.g., \cite{alencar12}).

% \subsubsection{Limits in distance and \Macc }
As shown in Sect.~\ref{low_res_mode}, observations of objects down to V$=$22\,mag with CUBES will be possible 
with the low-res. mode. By scaling the X-Shooter spectrum of Sz\,88\,A (V$=$13.2\,mag) to V$=$22\,mag, we derived 
line fluxes a factor $\sim$3$\times$10$^{-4}$ lower than those discussed in Sec~\ref{bal_decrs}. Thus, 
considering the {\em Gaia} DR2 distance of 158\,pc for Sz\,88\,A, and assuming \Av$=$0 mag, it would be possible in principle to observe strong accretors similar to Sz\,88\,A up to a distance of $\sim$9\,kpc with CUBES, 
with line fluxes in the range of 7.5$\times$10$^{-16}$ to 2.2$\times$10$^{-17}$ erg\,s$^{-1}$\,cm$^2$, 
respectively for the Balmer lines from \Ha\ to \ion{H}{15}, and down to 4.2$\times$10$^{-19}$ erg\,s$^{-1}$\,cm$^2$ 
for \ion{H}{25}.  Of course, the ultimate reach of such observations will be limited by extinction, particularly given the average empirical rule of one magnitude of extinction per kpc (implying \Av$\approx$9\,mag at 9\,kpc) combined with greater extinction towards the Galactic plane. Nonetheless, our tests highlight that CUBES will allow us to study YSOs at $\sim$kpc scale distances for the first time.

As seen in the previous sub-sections the high-resolution mode offers the possibility of observations of relatively 
reddened (\Av$=$3\,mag) YSOs with V$=$20\,mag (see Sect.~\ref{high_res_mode}), which translates to objects similar 
to Sz\,88\,A at a distance of up to $\sim$3.5\,kpc.

In the case of very low-mass objects (\Mstar$<$0.1\Msun), such as 2MASS\,J16095628$-$3859518 
(or Lup818s in \cite{alcala14} and \cite{alcala19}; SpT$=$M6; V$\approx$18\,mag; $\log$\Lacc$=-4.31$; 
$\log$\Macc$=-10.96$) in Lupus, the limiting V$=$22\,mag in the low-resolution mode implies a dilution
in flux by a factor $\sim$2.5$\times$10$^{-2}$, theoretically meaning a limiting $\log$(\Lacc)$\approx -6$ 
and $\log$(\Macc)~$\lesssim -11$, with \Lacc\ and \Macc\ in units of \Lsun\ and \Msun \,yr$^{-1}$, respectively.

\subsubsection{High Balmer lines and physical conditions of accreting gas}
\begin{figure}[!ht]
\resizebox{1\hsize}{!}{ {\includegraphics[bb=0 0 750 550]{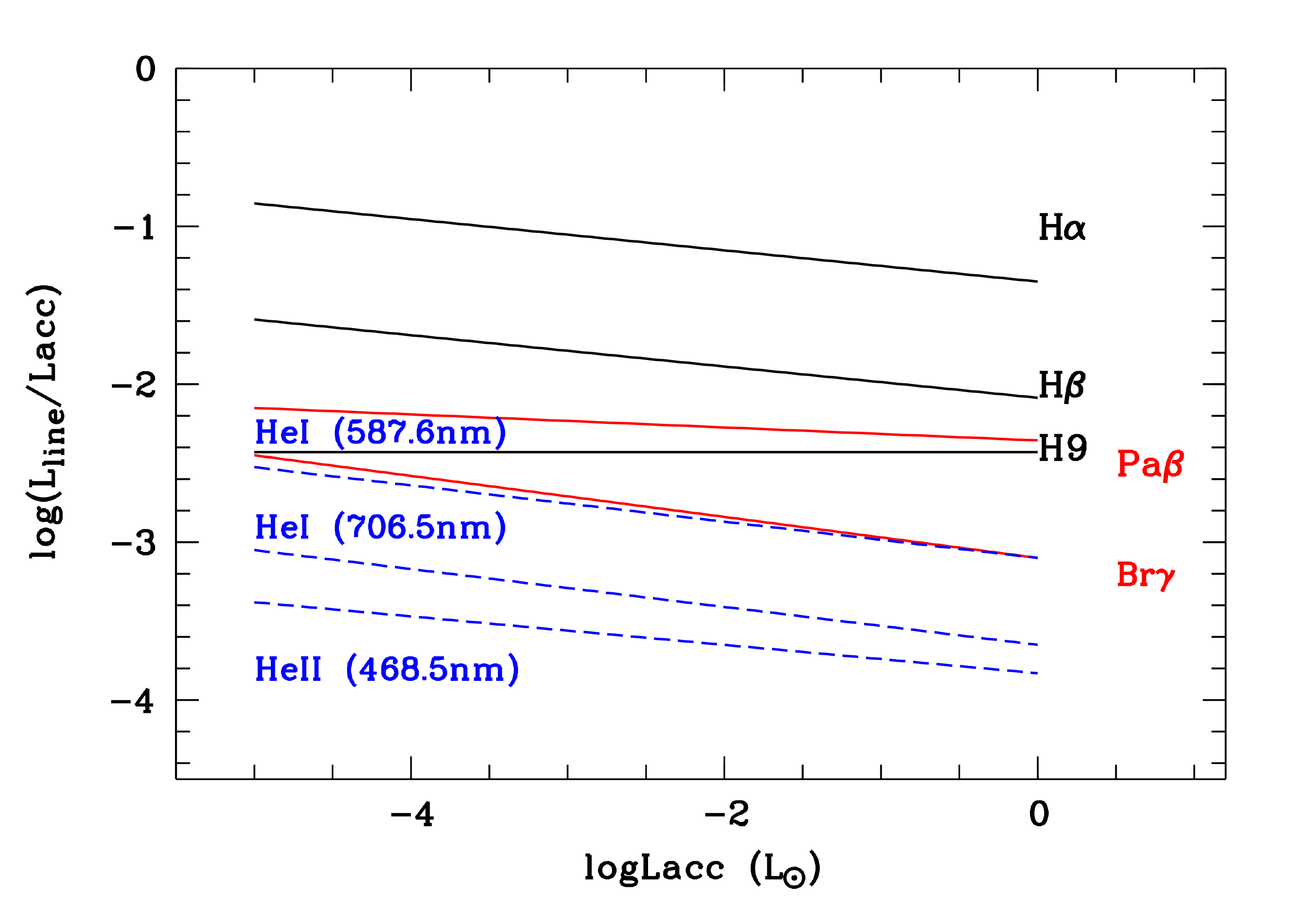}} }
\caption{Fractional \Ll/\Lacc\ line luminosity as a function  of \Lacc\ in log scale for example diagnostic
 lines as labelled. The relationships between line luminosity and accretion luminosity by \cite{alcala14}
 were used to produce the plot.
 \label{Ll_Lacc_vs_Lacc}}
\end{figure}

Our previous studies with X-Shooter (\cite{alcala14}, \cite{alcala17}, \cite{antoniucci17}) showed an important 
similarity of the physical conditions in the accreting gas for objects with very different \Macc, which was suggested
by a number of properties. One was the comparable behaviour of the Balmer decrements down to \ion{H}{15} for
different objects, while another was the similarity of the \Lacc--\Ll\ relations which, to zeroth order, are linear,
with slopes varying between 0.99 and 1.18 for the hydrogen lines, and 0.90 to 1.16 for the \ion{He}{I} lines. This
means that, over a range of five orders of magnitude in \Lacc, a similar fraction of the accretion energy is  emitted
in each line, independently of their excitation potential and optical depth. This is shown in 
Fig.~\ref{Ll_Lacc_vs_Lacc} where the fractional line luminosities (\Ll/\Lacc) for a few example lines are plotted as 
a function of \Lacc\ in log scale. We used the \Lacc--\Ll\ relationships by \cite{alcala14} to produce this plot. The
relations are very stable across the examined stellar parameters, suggesting that the physical conditions  of the
accreting gas are very similar in all objects. Some evidence of a different behaviour of the Balmer  decrements at
high Balmer lines, possibly indicating different physical conditions of the accreting gas,  were discussed in
Sect.~\ref{bal_decrs}.

%%%%%%%%%%%%%%%%%%%%%%%%%%%%%%% Fig_ %%%%%%%%%%%%%%%%%%%%%%%%%%%%%%%%%%%%%
%%%%%%%%%%%%%%%%%%%%%%%%%%%%%%%%%%%%%%%%%%%%%%%%%%%%%%%%%%%%%%%%%%%%%%%%%%%%

CUBES observations of relatively bright YSOs (with V$<$19\,mag) in the high-res. mode will allow us to investigate
the decrements in a more detailed fashion, as well as extending the \Lacc--\Ll\ calibrations  to the high-order
Balmer lines, hence further investigate the physical conditions of the accreting  gas at high quantum numbers. The
ODYSSEUS-PENELLOPE and GHOsT projects (see Sect.~\ref{Sect5})  will provide bright and very well characterised
targets (in terms of physical and accretion properties) that will be perfect targets for CUBES observations.  The
extended \Lacc--\Ll\ calibrations will be critical for estimates of \Lacc\ in much fainter objects,  where fits to
the Balmer jump will not be possible because of the low S/N of the continuum, but where the emission lines are
expected to be well detected.

CUBES alone would be in principle sufficient for these studies, but the simultaneous UVES link would provide 
more diagnostics in the red for a more complete analysis.
The resolution of CUBES will also be  helpful for the analysis of the line profiles, providing further
information on the physical and kinematical properties of the accreting gas (e.g. \cite{antoniucci17}).

\subsubsection{Accretion at advanced stages}
\label{advanced_stages}

Transition disks (TDs) are protoplanetary disks that show evidence of inner holes and gaps,  as observed in
millimeter interferometric observations (\cite{andrews11}, \cite{vandermarel18}, \cite{andrews18}) and in the dip of
the mid-IR spectral energy distribution (e.g. \cite{merin08}). The leading explanation for the presence of inner
holes and gaps in the dust distribution of TDs is  photo-evaporation (e.g.  \cite{ercolanopascucci17},
\cite{vandermarel18}).

Low-mass pre-main-sequence stars with transitional disks accreting at very low rates are likely in the final stages 
of inner disk evolution, and probably have already formed proto-planets (\cite{owen_clarke12}).  Hence, identifying
and investigating such slow accretors may help us to understand planet formation.  However, measurements of low
\Macc ~are challenging. In general, weak accretion is not easily detectable  in the region of the Balmer jump. Some
previous studies attempted \Macc ~measurements in TDs using other  tracers, such as trends between the lines width
and \Macc\  (e.g. \cite{merin10}, \cite{cieza10}).  However, such estimates depend heavily on rather uncertain
scaling relations (see discussion in \cite{alcala14}). Using the \Lacc--\Ll\ relationships is in principle possible.
However, at very low accretion levels the chromospheric contribution to the emission budget may be comparable to 
\Lacc\ and must therefore be  disentangled.

\begin{figure}[!ht]
\resizebox{1\hsize}{!}{ {\includegraphics[bb=0 0 750 550]{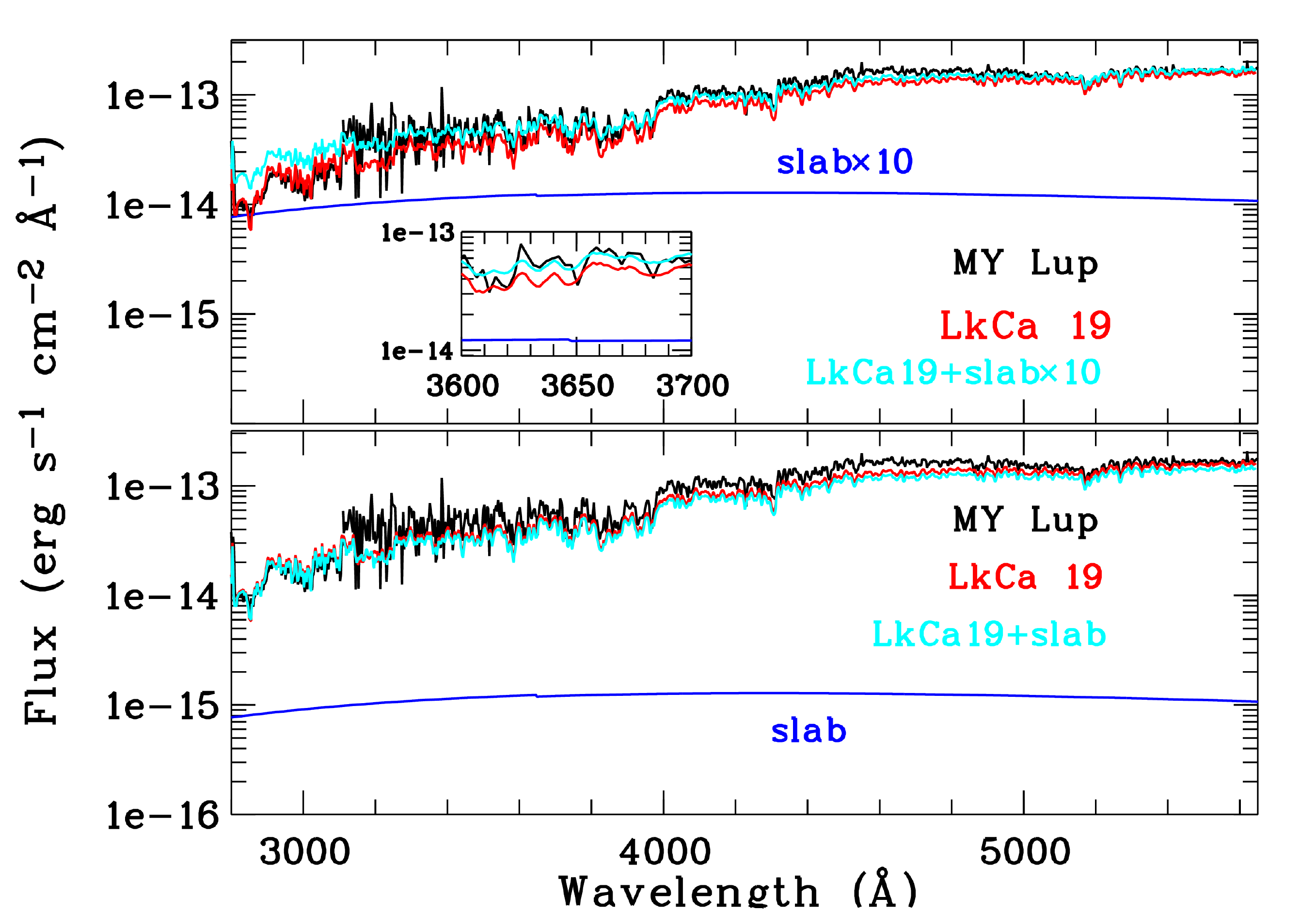}} }
\caption{Lower panel: extinction corrected low-resolution spectra of MY\,Lup (in black) 
and the non-accreting template LkCa\,19 (in red). The slab model derived in \cite{alcala17} 
is also shown (in blue). The cyan spectrum shows the sum of the non-accreting template plus 
the slab emission. Upper panel: as per the lower panel, but now with the slab model 
multiplied by ten. Note the tiny Balmer jump shown in the inset. Figure adapted from \cite{alcala19}.
 \label{MYLup_slab}}
 \end{figure}

Another important aspect regarding YSOs in advanced stages is represented by the i
possible degeneracy of stellar  and
accretion parameters in objects with spectral types earlier than about K3. In these relatively early-type objects
the low contrast between the continuum  excess  emission  and  the  photospheric + chromospheric  emission, may
prevent a reliable assessment of accretion.  For example, Fig.~\ref{MYLup_slab} shows the case of the star MY\,Lup;
the solution on the accretion parameters  based on low-res. spectra alone may be ambiguous, with up to an order of
magnitude difference in \Lacc\   (see e.g. \cite{alcala19}). CUBES will allow us to measure low accretion rates, by
more precisely disentangling the contribution from chromospheric emission, which is difficult to remove 
in general, but practically impossible if S/N is low in the NUV. The simultaneous link with UVES 
will improve the fitting of continuum excess emission.

%%%%%%%%%%%%%%%%%%%%%%%%%%%%%%% Fig_ %%%%%%%%%%%%%%%%%%%%%%%%%%%%%%%%%%%%%
%%%%%%%%%%%%%%%%%%%%%%%%%%%%%%%%%%%%%%%%%%%%%%%%%%%%%%%%%%%%%%%%%%%%%%%%%%%%

\subsubsection{Low end of \Macc\ in large surveys}
\label{low_end_of_Macc}

% \paragraph{\underline{\Mstar, \Macc\ and \Mdisk\ scaling relationships}}
A crucial prediction of the YSO viscous accretion theory is the existence of correlations between the disk mass
(\Mdisk) the stellar mass (\Mstar) and the mass accretion rate (\Macc) (\cite{hartmann16}). Such correlations have 
been confirmed only recently for the disk populations of the 1-3\,Myr old Lupus and Chamaeleon\,I star-forming 
regions (see Sect.~\ref{disk_surveys} and \cite{manara16} and references therein). The observed relations show, 
however, a much larger scatter in \Macc\ ($>$0.9\,dex) than viscous evolution would predict for 1-3\,Myr old
populations.  The viscous disk theory  predicts a smaller spread of \Macc\ ($<$0.4\,dex) at older ages. We thus need
to measure  the spread of \Macc\ at given \Mdisk\ for carefully selected samples of relatively old YSOs, where
accretion activity is  expected to be weak. Recent work with X-Shooter showed the existence of very high accretion
rates in the ($>$5\,Myr)  Upper Scorpius association. CUBES will contribute by measuring the weak Balmer jump and
weak excess emission, as well as disentangling the contribution of the chromospheric emission.

\begin{figure}[!ht]
\resizebox{1.0\hsize}{!}{ {\includegraphics[bb=5 0 52 40]{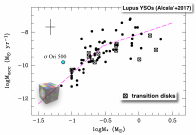}} }
\caption{Mass accretion rate (\Macc) as a function of stellar mass (\Mstar) for YSOs in the Lupus clouds (black symbols).  
Transitional disks are shown with crossed squares. The position of the young brown dwarf 2MASS\,J05382543$-$0242412 
(or $\sigma$-Ori\,500) is shown with the blue dot.
 Average errors in $\log$(\Macc) and $\log$(\Mstar) are shown in the upper left. 
 Dashed magenta lines represent the fits to the data as in Eqs. (4) and (5) from 
 \cite{alcala17}. The CUBES logo shows the region of the diagram that the instrument 
 may be able to reach. Figure adapted from \cite{alcala19}.
\label{Macc_Mstar_plot}}
\end{figure}

Another important point which deserves further investigation is that there is growing evidence 
of a change in the slope of the \Mstar--\Macc\  relationship for CTTS with ages of 
2-3\,Myr at \Mstar$<$0.2\,\Msun\ (\cite{manara17} and \cite{alcala17}, and see Fig.~\ref{Macc_Mstar_plot}).
Such a break could be related to a faster disk evolution at low-masses (e.g. \cite{VorovyobBasu09}). 
To verify this, the slope of the \Macc--\Mstar\ relationship needs to be sampled at much lower \Mstar\ and \Macc\ 
values than done so far. In Fig.~\ref{Macc_Mstar_plot} the  theoretical limits that CUBES may be able to 
reach are shown with the instrument logo. Note that such limits are much lower than the values for 
$\sigma$-Ori\,500 (2MASS\,J05382543$-$0242412), the V$=$ 20\,mag young brown dwarf discussed in Sect.~\ref{bal_decrs}.

%%%%%%%%%%%%%%%%%%%%%%%%%%%%%%% Fig_ %%%%%%%%%%%%%%%%%%%%%%%%%%%%%%%%%%%%%
%%%%%%%%%%%%%%%%%%%%%%%%%%%%%%%%%%%%%%%%%%%%%%%%%%%%%%%%%%%%%%%%%%%%%%%%%%%%

CUBES not only will complement the surveys of 1-2\,Myr clusters with accurate \Macc\ determinations, 
but its sensitivity will allow us to investigate the accretion process in young brown dwarfs, 
possibly down to \Mstar$\sim$30\Mjup\  and  \Macc\ $\lesssim 10^{-11}$\,\Msun/yr.
By studying the \Macc--\Mstar\ relationship at such limits it will be possible to 
investigate whether low-mass objects have indeed a faster disk evolution than their more
massive counterparts, as proposed by some disk accretion models (\cite{VorovyobBasu09}).

Finally, another important issue regards the validity of the \Lacc--\Ll\ relationships 
at the very low-mass regime. Recently \cite{aoyama21} have produced a theoretical \Lacc -- L(\Ha) relationship drawn 
from modelling of a shock at the surface of a planetary mass object. These authors show that the new theoretical 
relationship yields a much higher \Lacc\ for a given L(\Ha) than what the relationships drawn from YSO spectra 
would predict. Hence, the validation of the relationships at the lowest possible mass is of key importance. 

The high-sensitivity makes of CUBES a self-standing instrument for the studies of accretion in 
very-low mass YSOs, although the simultaneous link with UVES would be helpful for the YSO classification.

\subsubsection{\Macc\ in low metallicity environments}
A very important aspect that has received little attention so far, mainly because of the limited 
sensitivity of the existing intermediate/high resolution instruments, is the spectroscopic study 
of YSOs in distant  (a few kpc) star-forming regions. At such distances, the low metallicity 
may have an important impact on accretion. 

Photometric studies in low-metallicity regions (e.g. \cite{demarchi17}, \cite{biazzo19} and 
references therein) suggest that metal-poor stars accrete at higher rates than solar-metallicity 
stars in nearby Galactic star-forming regions. As a consequence, most stars forming in a 
low-metallicity environment should experience disk dispersal at an earlier stage (in less 
than about 1\,Myr) than those forming in solar metallicity environments (where disk dispersal 
may last up to 5-6\,Myr). Thus, metallicity may play a crucial role on circumstellar disk 
evolution, hence on the time scale available for planet formation.

Typical solar-mass YSOs at kpc distances are faint (V$>$18\,mag, see \cite{cusano11}) so require 
the significant increase in sensitivity that CUBES will provide compared to current instruments.
CUBES observations will allow us to investigate the Balmer jump, hence \Lacc\ and \Macc\ at 
low Z ($<$0.2\,Z$_{\odot}$) and answer questions such as:  
are the \Lacc--\Lstar\ relationships the same at Solar metallicity? Are the Balmer 
decrements similar to those at Solar metallicity? What are the effects of local and external 
UV radiation fields (i.e. external photo-evaporation)?

CUBES will enable studies of accretion at low metallicities, as well as the effects of 
local UV fields on accretion and winds -- weak fields should not influence the disks, 
whereas strong fields may modify the ionisation rate in the disk leading to larger 
mass-loss rates (\cite{guarcello21}). 

 CUBES can be a self-standing instrument for this science case, but YSO crowding may represent 
an issue for the farthermost SFRs.

\subsection{Winds and outflows}
As mentioned in Sect. 2.3, the outflow activity in YSOs is a phenomenon strictly related to accretion  at different 
mass regimes. The study of such outflows enables us to elucidate the relationship between mass  ejection and accretion,
 and test the universality of the star formation mechanism as a function of central  YSO mass.

In particular, forbidden lines from the UV to the NIR spectral domain can be used to measure the physical properties 
(density, temperature and ionisation fraction) of the different outflow manifestations    using diagnostic tools
specifically  designed for the purpose. Analysis combining lines at different wavelengths originating  from various
excitation regimes is important, as applied to CTTS jets observed with X-Shooter (\cite{bacciotti11},
\cite{giannini15a}) and more recently with the GIARPS instrument  (\cite{giannini19}) at much higher resolution. The
CUBES + UVES combination will provide sufficient sensitivity and  resolution to apply such diagnostic tools to 
studies of fainter and more distant YSOs jets than possible so far.

To this aim, the CUBES spectral range covers several key lines for the study of jets/winds in YSOs. We already 
mentioned the importance of the [\ion{O}{ii}] 3726\AA\ line, in combination with the [\ion{O}{i}] 6300\,\AA\  line, to
estimate the ionization degree of the jets/winds plasma. Other important lines providing complementary information or 
tracing excitation regimes not probed by optical/IR lines are also present in the CUBES + UVES range. For example,  
the [\ion{S}{ii}] doublet at 4069/4076\,\AA\ is a useful diagnostic of dense winds in protoplanetary disks 
(\cite{ballabio20} and references therein). Also, the little-studied  [\ion{Ne}{iii}] 3869\,\AA\ line signals the
presence of high-energy photons, thus probing either high velocity  ($>$ 100 km/s) shocks or X-ray heating
\cite{liu14,liu21}. The two scenarios can potentially be discriminated through  CUBES + UVES velocity-resolved
observations.

Finally, the NUV spectral range includes several bright [\ion{Fe}{ii}] and [\ion{Fe}{iii}] lines produced  from high
energy levels, which are therefore good tracers of the temperature of the ionised gas. The observation of 
[\ion{Fe}{iii}], combined with [\ion{Fe}{ii}], allows the simultaneous determination of temperature, density and degree
of ionisation, as well as a precise estimate of the Fe abundance in the gas. The relevant non-LTE models required to
predict the line fuxes already exist (e.g., \cite{giannini15}, \cite{giannini15a}).

This analysis approach can be applied to a large sample of sources having different stellar and  accretion properties.
In addition, the CUBES spectral resolution will enable us to  distinguish the different  kinematical wind components 
and separately study the properties of the jets and of the slow winds, as well  as investigating whether the relative
importance of these two outflow components changes with age, when the  contribution of photo-evaporation is expected 
to be dominant.

Following determination of the physical parameters, the mass ejection rate, \Mout, can be derived from the luminosity 
of spectrally-resolved forbidden  lines. This will allow us to measure the  \Mout/\Macc\, parameter, which critically
depends on the jet launching mechanism, and in turn to infer possible variations in the mass loss efficency due to
modifications of the inner disk magnetic and physical properties.

While longer-wavelength data are required for this case, it is not a strong requirement that they are simultaneous in time. In this example, the benefit of the UVES fibrelink is primarily more efficient operations than separate observations.

\section{Summary and conclusions}
\label{Sect6}

The CUBES science case on accretion and outflows in YSOs has been presented. The performance of CUBES,  mainly in terms of its increased sensitivity in the NUV (3000--4050\,\AA), will allow us to broaden
investigations of the evolution of circumstellar disks,  by studying the accretion, jets/winds and 
photo-evaporation processes down to the low-mass, brown dwarf regime and in more distant YSOs than 
done so far. In particular CUBES will enable:

\begin{itemize}

\item more precise flux measurements of continuum and emission lines in the NUV.  Contemporaneous 
photometric observations in the same spectral range will be needed in case the CUBES flux calibration 
will not be precise at the 15\% level or better.

\item detection of weak emission lines (and their line morphology), in particular for the 
\ion{Ca}{ii} H\&K and high-H lines, which are diagnostics of the accretion funnel flows.

\item investigation of the physical conditions of the in-flowing gas via precise  Balmer decrement determinations,
including the high-order Balmer lines.

\item refinement of the \Lacc--\Ll\ relationships by including Balmer lines with a high quantum number.

\item better definition of the Balmer jump, as well as detection of weak emission lines, allowing for studies of
low-accretion YSOs, including objects in the latest stages of evolution.

\item studies of accretion and outflows in very low-mass browns dwarfs, and to study YSOs at much larger distances
than done so far, enabling investigation of the effects of metallicity and external UV-radiation on accretion and
winds.

\end{itemize}
 
In conclusion, CUBES observations of YSOs will lead to a step-change in the study of accretion and outflows
in solar-type young stars hence, on protoplanetary disk evolution, as the spectrograph will enable investigation 
of targets that are several magnitudes fainter than those reachable with current instrumentation.

\begin{acknowledgements}
 We very much thank the two anonymous referees for their comments and suggestions.
Financial support from the project  PRIN  INAF  2019 ''Spectroscopically tracing the disk dispersal 
evolution (STRADE)" is warmly acknowledged. This research made use of the SIMBAD database, operated 
at the CDS (Strasbourg, France). 
\end{acknowledgements}

% Authors must disklose all relationships or interests that 



\end{document}